\documentclass[11pt]{article}
\usepackage[english]{babel}
\usepackage[latin1]{inputenc}
\usepackage{amsmath,amssymb,indentfirst}
\usepackage[dvips]{graphics}
\usepackage{epsfig}

\newcommand{\tetn}{{\mbox{\boldmath $\theta$}}}

\newcommand{\omegn}{{\mbox{\boldmath $\omega$}}}

\newcommand{\taun}{{\mbox{\boldmath $\tau$}}}

\title{\textbf{A New Generalized Kumaraswamy Distribution}}
\author
{Jalmar M. F. Carrasco
\footnote
{E-mail: jalmar@ime.usp.br} \\
Departamento de Estatística\\
Universidade de São Paulo\\
Brazil
\and
Silvia L. P. Ferrari
\footnote
{Corresponding author: Departamento de Estatística, Universidade de São Paulo,
Rua do Matão, 1010, 05508-090,  São Paulo, SP, Brazil.
E-mail: silviaferrari.usp@gmail.com} \\
Departamento de Estatística\\
Universidade de São Paulo\\
Brazil
\and
Gauss M. Cordeiro.
\footnote{E-mail: gausscordeiro@uol.com.br} \\
Departamento de Estatística e Informática\\
Universidade Federal Rural de Pernambuco\\
Brazil}
\date{}

\begin{document}

\maketitle


\begin{abstract}
A new five-parameter continuous distribution which generalizes the Kumaraswamy
and the beta distributions as well as some other well-known distributions is proposed
and studied. The model has as special cases new four- and three-parameter distributions
on the standard unit interval. Moments, mean deviations, R\'enyi's
entropy and the moments of order statistics are obtained for the new 
generalized Kumaraswamy distribution. The score function is given and estimation is performed by maximum likelihood. Hypothesis testing is also discussed. A data set is used to illustrate an application of the proposed distribution.

\vspace{5mm} \noindent {\em Keywords}: Beta distribution; Continuous proportions; Generalized Kumaraswamy distribution; Kumaraswamy distribution;  Maximum likelihood; McDonald Distribution; Moments.

\end{abstract}

\section{Introduction}

We introduce a new five-parameter distribution, so-called generalized
Kumaraswamy (\textit{GKw}) distribution, which contains some well-known distributions
as special sub-models as, for example, the Kumaraswamy (\textit{Kw}) and beta
($\mathfrak{B}$) distributions. The \textit{GKw} distribution allows us to define new
three- and four-parameter generalizations of such distributions. The new model can be
used in a variety of problems for modeling continuous proportions data due to its
flexibility in accommodating different forms of density functions. 

The \textit{GKw} distribution comes from the following idea. Wahed (2006) and Ferreira and Steel (2006)
demonstrated that any parametric family of distributions can be incorporated into larger
families through an application of the probability integral transform. Specifically,
let $G_1(\cdot;\omegn)$ be a cumulative distribution function (cdf) with corresponding
probability density function (pdf) $g_1(\cdot;\omegn)$, and $g_2(\cdot;\taun)$ be a pdf
having support on the standard unit interval. Here, $\omegn$ and $\taun$ represent
scalar or vector parameters. Now let
\begin{eqnarray}\label{F}
F(x;\omegn,\taun)=\int_{0}^{G_1(x;\omegn)} g_2(t;\taun) dt.
\end{eqnarray}
Note that $F(\cdot;\omegn,\taun)$ is a cdf and that $F(\cdot;\omegn,\taun)$ and 
$G_1(x;\omegn)$ have the same support. The pdf corresponding to (\ref{F}) is
\begin{eqnarray}
\label{f}
f(x;\omegn,\taun)=g_{2}(G_{1}(x;\omegn);\taun)g_{1}(x;\omegn).
\end{eqnarray}
This mechanism for defining generalized distributions from a parametric
cdf $G_1(\cdot;\omegn)$ is particularly attractive when $G_1(\cdot;\omegn)$
has a closed-form expression.

The beta density is often used in place of $g_{2}(\cdot;\taun)$. However,
different choices for $G_1(\cdot;\omegn)$ have been considered in the
literature. Eugene et al. (2002) defined the beta normal distribution
by taking $G_1(\cdot;\omegn)$ to be the cdf of the standard normal distribution and derived
some of its first moments. More general expressions for these moments were obtained
by Gupta and Nadarajah (2004a). Nadarajah and Kotz (2004) introduced the beta Gumbel (\textit{BG}) distribution by taking $G_1(\cdot;\omegn)$ to be the cdf of the Gumbel distribution and provided
closed form expressions for the moments, the asymptotic distribution of the extreme order statistics
and discussed the maximum likelihood estimation procedure. Nadarajah and Gupta (2004)
introduced the beta Fréchet (\textit{BF}) distribution by taking $G_1(\cdot;\omegn)$ to be
the Fréchet distribution, derived the analytical shapes of its density and hazard rate functions
and calculated the asymptotic distribution of its extreme order statistics.
Also, Nadarajah and Kotz (2006) dealt with the beta exponential (\textit{BE}) distribution
and obtained its moment generating function, its first four cumulants, the asymptotic
distribution of its extreme order statistics and discussed maximum likelihood estimation.

The starting point of our proposal is the Kumaraswamy (\textit{Kw})
distribution (Kumaraswamy, 1980; see also Jones, 2009). It is very similar to the beta
distribution but has a closed-form cdf given by
\begin{eqnarray}\label{G1}
G_{1}(x;\omegn)&=&1-(1-x^\alpha)^{\beta},\,\,\,0<x<1,
\end{eqnarray}
where $\omegn=(\alpha,\beta)^\top$, $\alpha>0$ and $\beta>0$.
Its pdf becomes
\begin{eqnarray}\label{g1}
g_{1}(x;\omegn)&=&\alpha \beta x^{\alpha-1}(1-x^\alpha)^{\beta-1},\,\,0<x<1.
\end{eqnarray}
If $X$ is a random variable with pdf (\ref{g1}), we write $X \sim {\rm Kw}(\alpha,\beta)$.
The \textit{Kw} distribution was originally conceived to model hydrological
phenomena and has been used for this and also for other purposes. See, for example,
Sundar and Subbiah (1989), Fletcher and Ponnambalam (1996), Seifi et al. (2000),
Ganji et al. (2006), Sanchez et al. (2007) and Courard-Hauri (2007).

In the present paper, we propose a generalization of the \textit{Kw} distribution by
taking $G_1(\cdot;\omegn)$ as cdf (\ref{G1}) and $g_2(\cdot;\taun)$ as the standard
generalized beta density of first kind (McDonald, 1984), with pdf given by
\begin{eqnarray}
\label{g2}
g_2(x;\taun)=\frac{\lambda x^{\lambda \gamma-1}(1-x^{\lambda})^{\eta-1}}{B(\gamma,\eta)},\,\,0<x<1,
\end{eqnarray}
where $\taun=(\gamma,\eta,\lambda)^\top$, $\gamma>0,\eta>0$ and $\lambda>0$,
$B(\gamma,\eta)=\Gamma(\gamma)\Gamma(\eta)/\Gamma(\gamma+\eta)$ is the beta
function and $\Gamma(\cdot)$ is the gamma function. If $X$ is a random variable with
density function (\ref{g2}), we write $X\sim {\rm GB1}(\gamma,\eta,\lambda)$.
Note that if $X\sim {\rm GB1}(\gamma,\eta,1)$ then $X\sim  \mathfrak{B}(\gamma,\eta)$,
i.e., $X$ has a beta distribution with parameters $\gamma$ and $\eta$.

The article is organized as follows. In Section 2, we define the
\textit{GKw} distribution, plot its density function for selected parameter values
and provide some of its mathematical properties. In Section 3, we present some special
sub-models. In Section 4, we obtain expansions for the distribution and density
functions. We demonstrate that the \textit{GKw} density can be expressed as a mixture
of \textit{Kw} and power densities. In Section 5, we give general formulae for the
moments and the moment generating function. Section 6 provides an expansion for the
quantile function. Section 7 is devoted to mean deviations about the mean and the median
and Bonferroni and Lorenz curves. In Section 8, we derive the density function of the
order statistics and their moments. The R\'enyi entropy is calculated in Section 9.
In Section 10, we discuss maximum likelihood estimation and determine the elements of
the observed information matrix. Section 11 provides an application to a real data set.
Section 12 ends the paper with some conclusions.

\section{The New Distribution}

We obtain an appropriate generalization of the \textit{Kw} distribution
by taking $G_{1}(\cdot;\omegn)$ as the two-parameter \textit{Kw} cdf (\ref{G1}) and associated pdf (\ref{g1}). For $g_{2}(\cdot;\taun)$, we consider a three-parameter generalized beta density of first kind given by (\ref{g2}). To avoid non-identifiability problems, we allow $\eta$ to vary
on $[1,\infty)$ only. We then write $\delta=\eta-1$ which varies on $(0,\infty)$.
Using (\ref{F}), the cdf of the \textit{GKw} distribution, with five positive
parameters $\alpha$, $\beta$, $\gamma$, $\delta$ and $\lambda$, is defined by
\begin{eqnarray}\label{FGKw}
F(x;\tetn)=\frac{\lambda}{B(\gamma,\delta+1)}\int_{0}^{1-(1-x^\alpha)^{\beta}}y^{\gamma \lambda-1}
(1-y^{\lambda})^{\delta}dy,
\end{eqnarray}
where $\tetn=(\alpha,\beta,\gamma,\delta,\lambda)^{\top}$ is the parameter vector.

The pdf corresponding to (\ref{FGKw}) is straightforwardly obtained from (\ref{f}) as
\begin{eqnarray}\label{fGKw}
f(x;\tetn)=\frac{\lambda\alpha\beta
x^{\alpha-1}}{B(\gamma,\delta+1)}(1-x^{\alpha})^{\beta-1}[1-(1-x^{\alpha})^{\beta}]^{\gamma \lambda-1}
\{1-[1-(1-x^{\alpha})^{\beta}]^{\lambda}\}^{\delta},\,\,0<x<1.
\end{eqnarray}
Based on the above construction, the new distribution can also be referred
to as the McDonald Kumaraswamy (\textit{McKw}) distribution. If $X$ is a random variable with
density function (\ref{fGKw}), we write $X \sim {\rm GKw}(\alpha,\beta,\gamma,\delta,\lambda)$.

An alternative, but related, motivation for (\ref{FGKw}) comes through the beta
construction (Eugene et al., 2002). We can easily show that
\begin{eqnarray}\label{FGKw1}
F(x;\tetn)=I_{[1-(1-x^\alpha)^{\beta}]^{\lambda}}(\gamma,\delta+1),
\end{eqnarray}
where $I_{x}(a,b)=B(a,b)^{-1}\int_0^{x}\omega^{a-1}(1-\omega)^{b-1} d\omega$
denotes the incomplete beta function ratio. Thus, the \textit{GKw} distribution
can arise by taking the beta construction applied to a new distribution,
namely the exponentiated Kumaraswamy (\textit{EKw}) distribution, to yield
(\ref{fGKw}), which can also be called the beta exponentiated Kumaraswamy (\textit{BEKw})
distribution, i.e., a beta type distribution defined by the baseline cumulative
function $G(x)=[1-(1-x^\alpha)^{\beta}]^{\lambda}$.

\begin{figure}
\begin{minipage}[b]{0.30\linewidth}
\includegraphics[width=\linewidth]{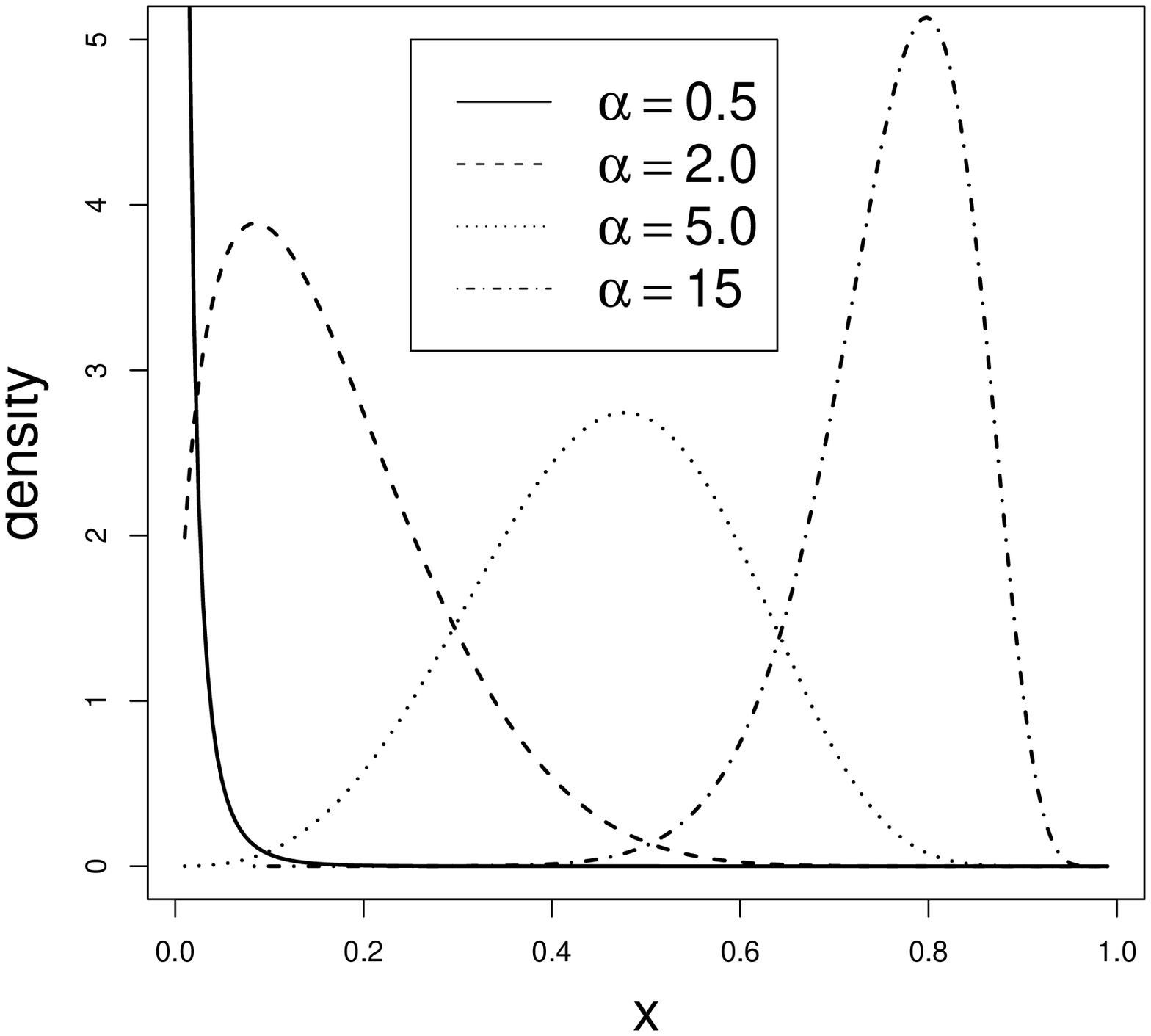}
\begin{center}
(a)
\end{center}
\end{minipage} \hfill
\begin{minipage}[b]{0.30\linewidth}
\includegraphics[width=\linewidth]{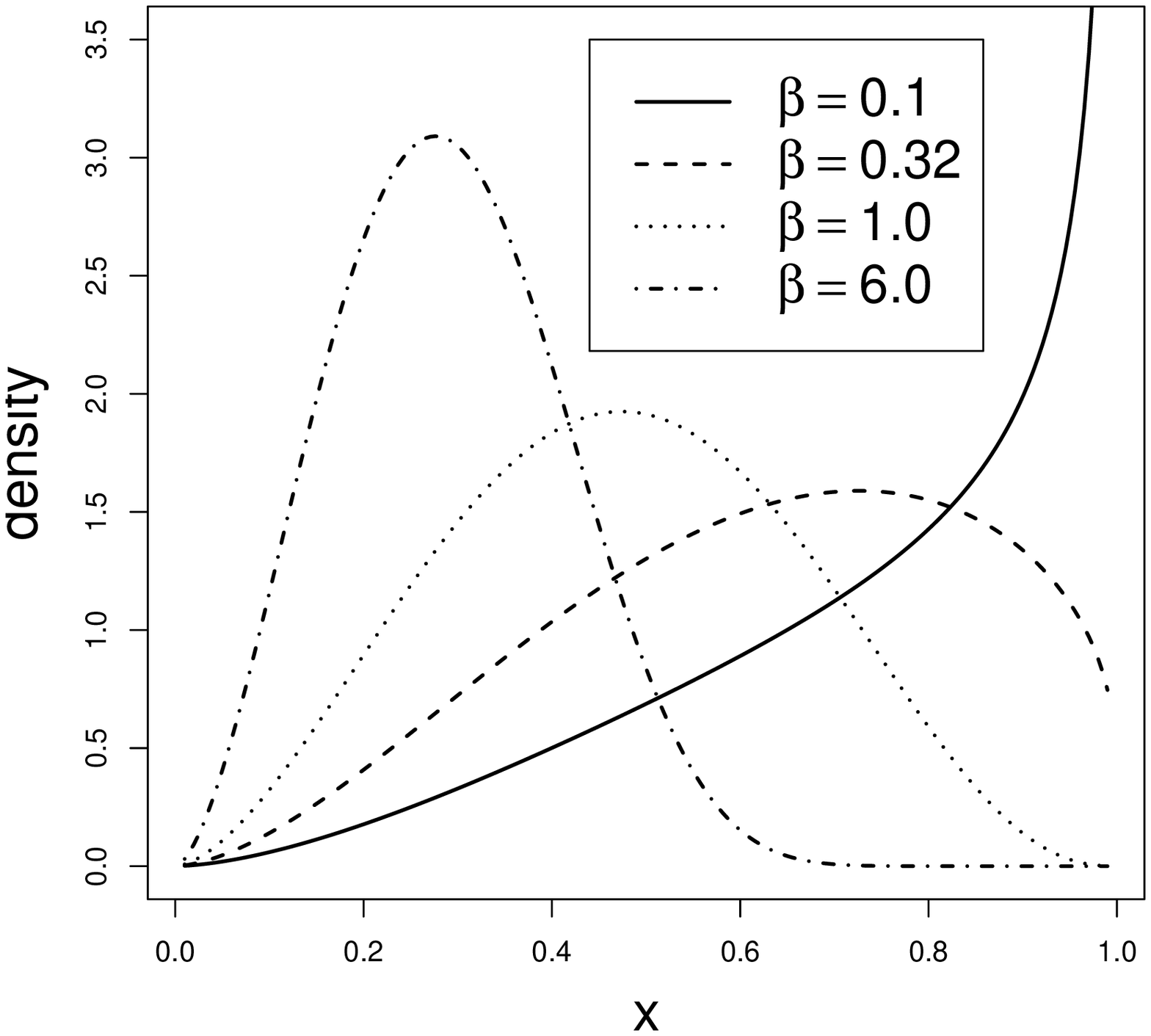}
\begin{center}
(b)
\end{center}
\end{minipage}\hfill
\begin{minipage}[b]{0.30\linewidth}
\includegraphics[width=\linewidth]{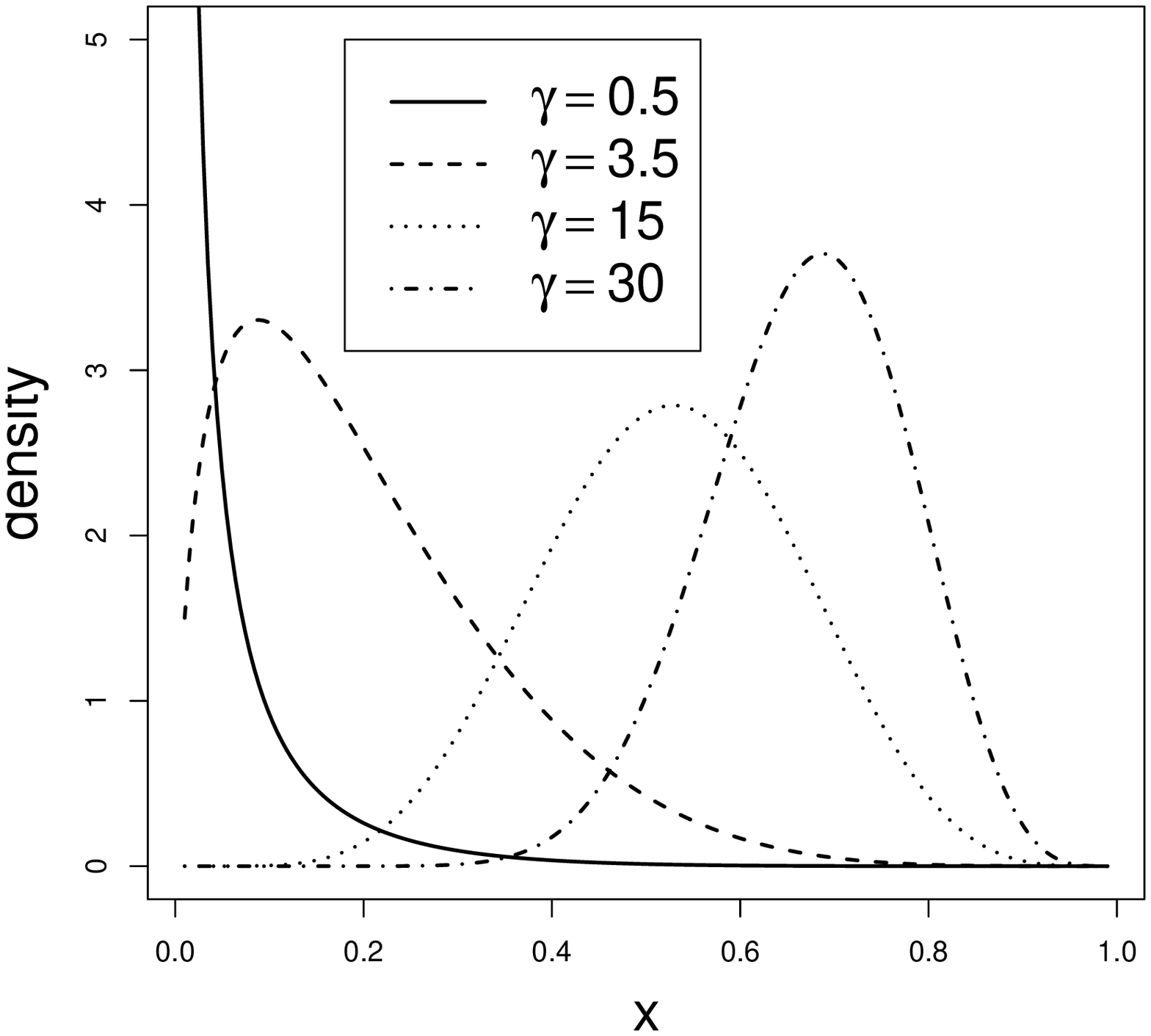}
\begin{center}
(c)
\end{center}
\end{minipage}
\\
\begin{minipage}[b]{0.30\linewidth}
\includegraphics[width=\linewidth]{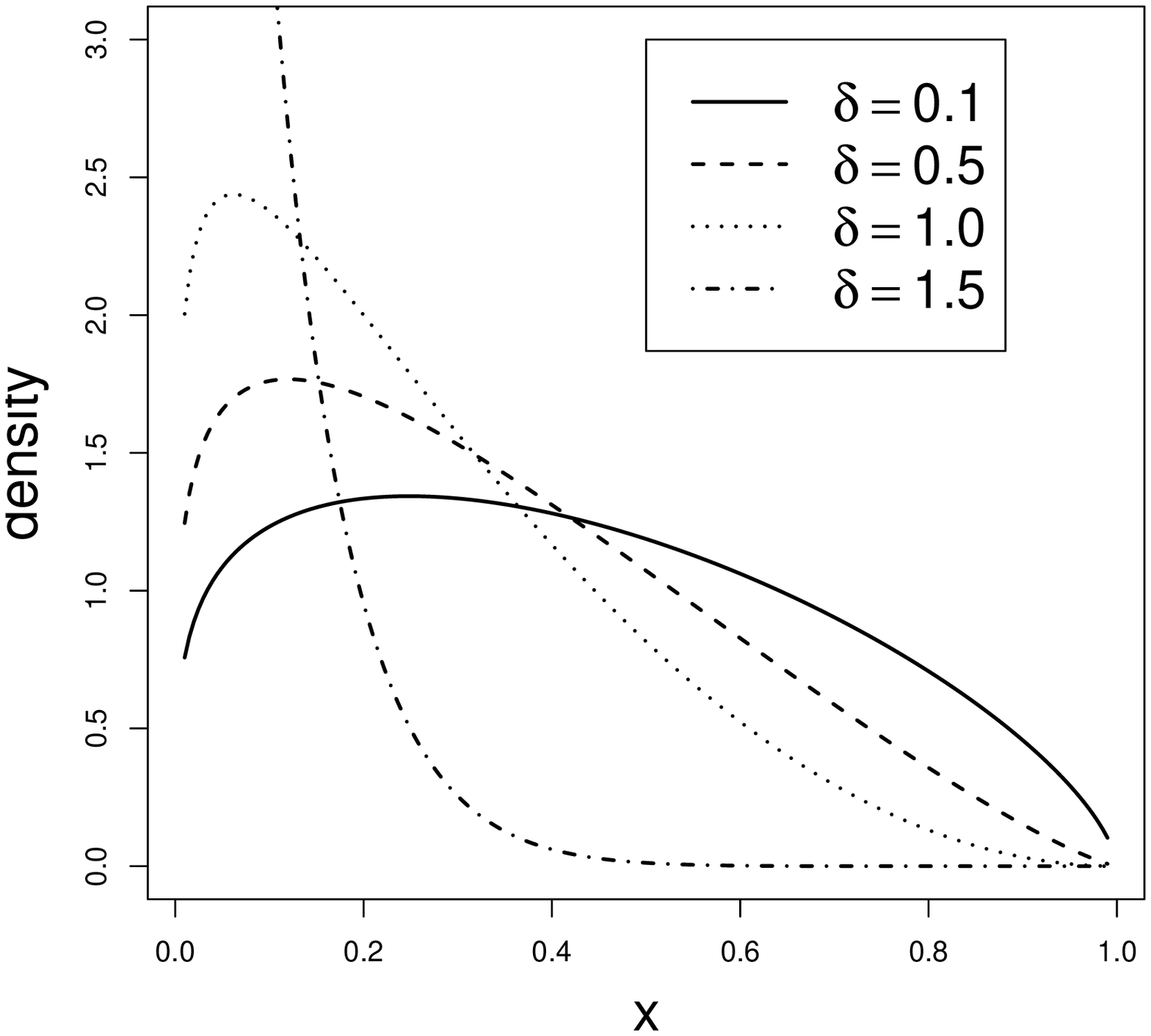}
\begin{center}
(d)
\end{center}
\end{minipage} \hfill
\begin{minipage}[b]{0.30\linewidth}
\includegraphics[width=\linewidth]{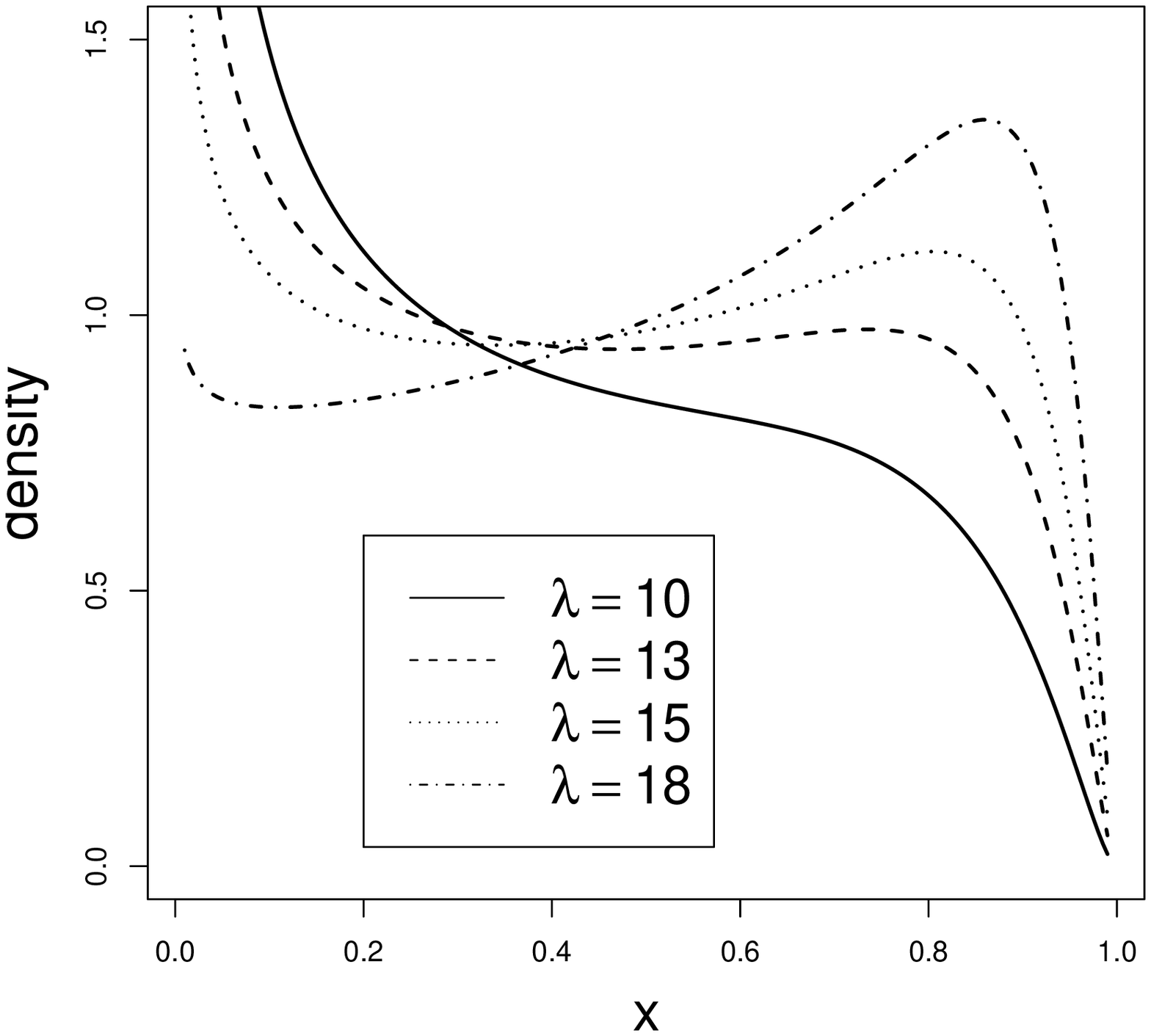}
\begin{center}
(e)
\end{center}
\end{minipage}\hfill
\begin{minipage}[b]{0.30\linewidth}
\includegraphics[width=\linewidth]{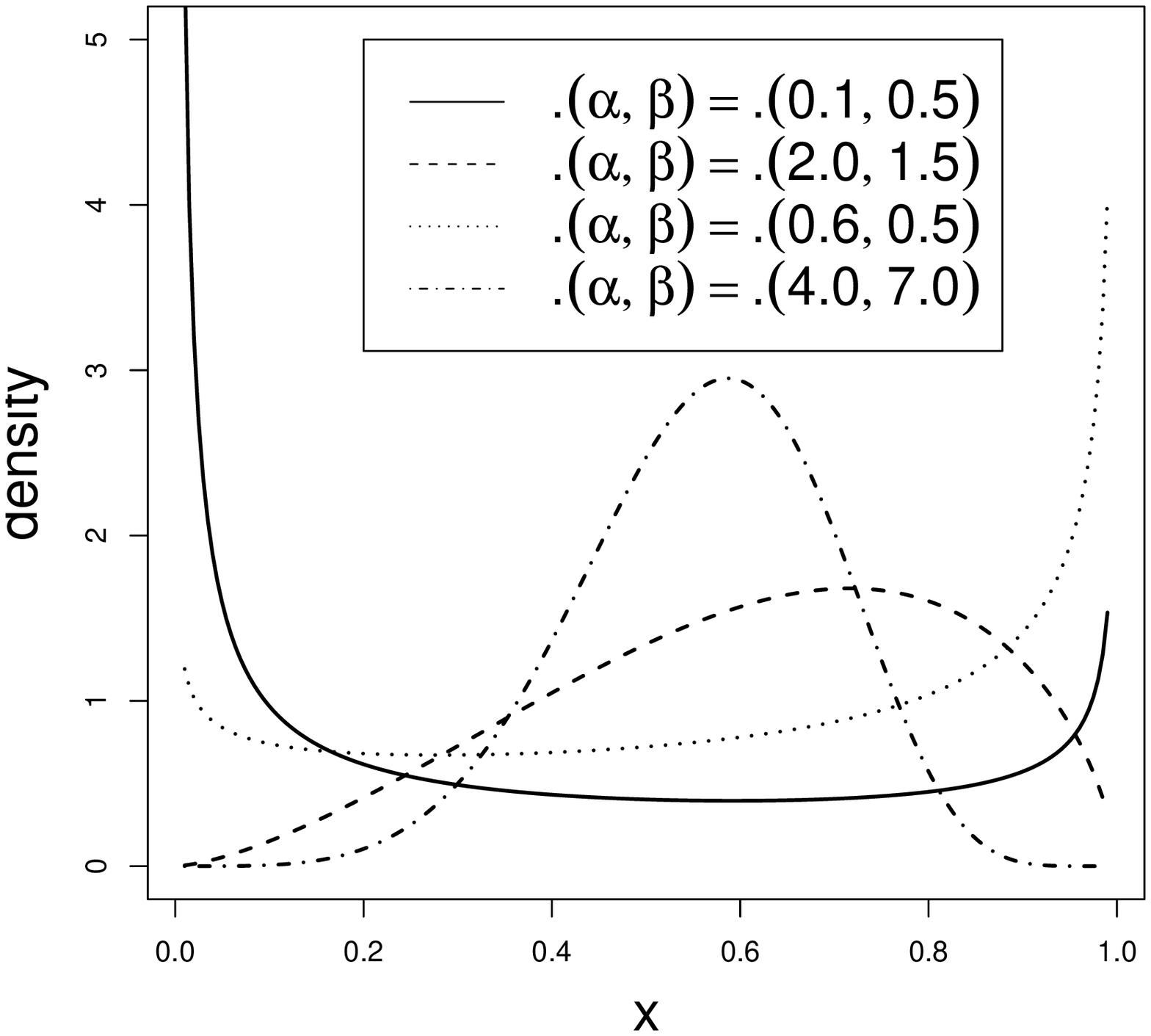}
\begin{center}
(f)
\end{center}
\end{minipage}
\\
\begin{minipage}[b]{0.30\linewidth}
\includegraphics[width=\linewidth]{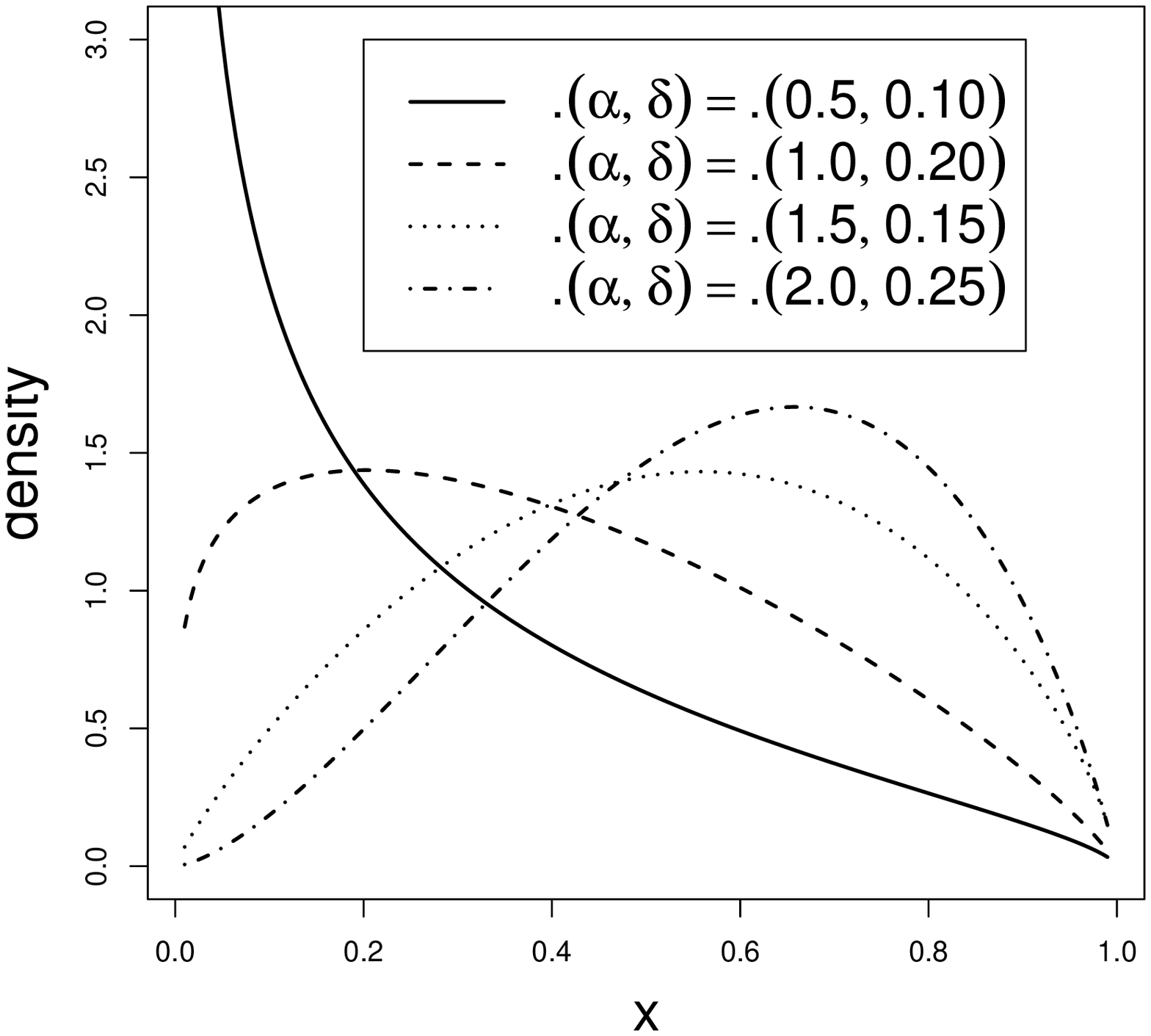}
\begin{center}
(g)
\end{center}
\end{minipage} \hfill
\begin{minipage}[b]{0.30\linewidth}
\includegraphics[width=\linewidth]{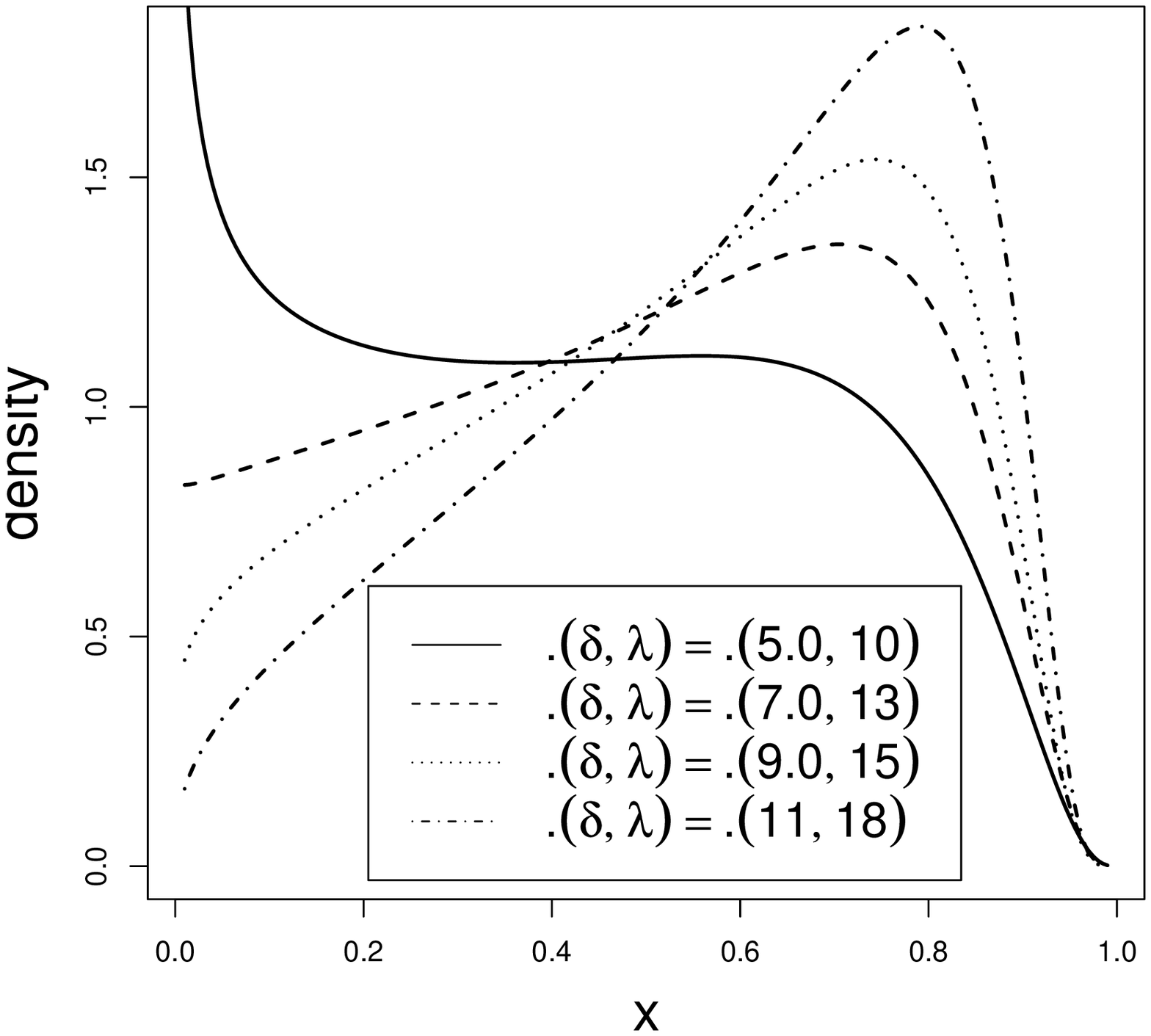}
\begin{center}
(h)
\end{center}
\end{minipage}\hfill
\begin{minipage}[b]{0.30\linewidth}
\includegraphics[width=\linewidth]{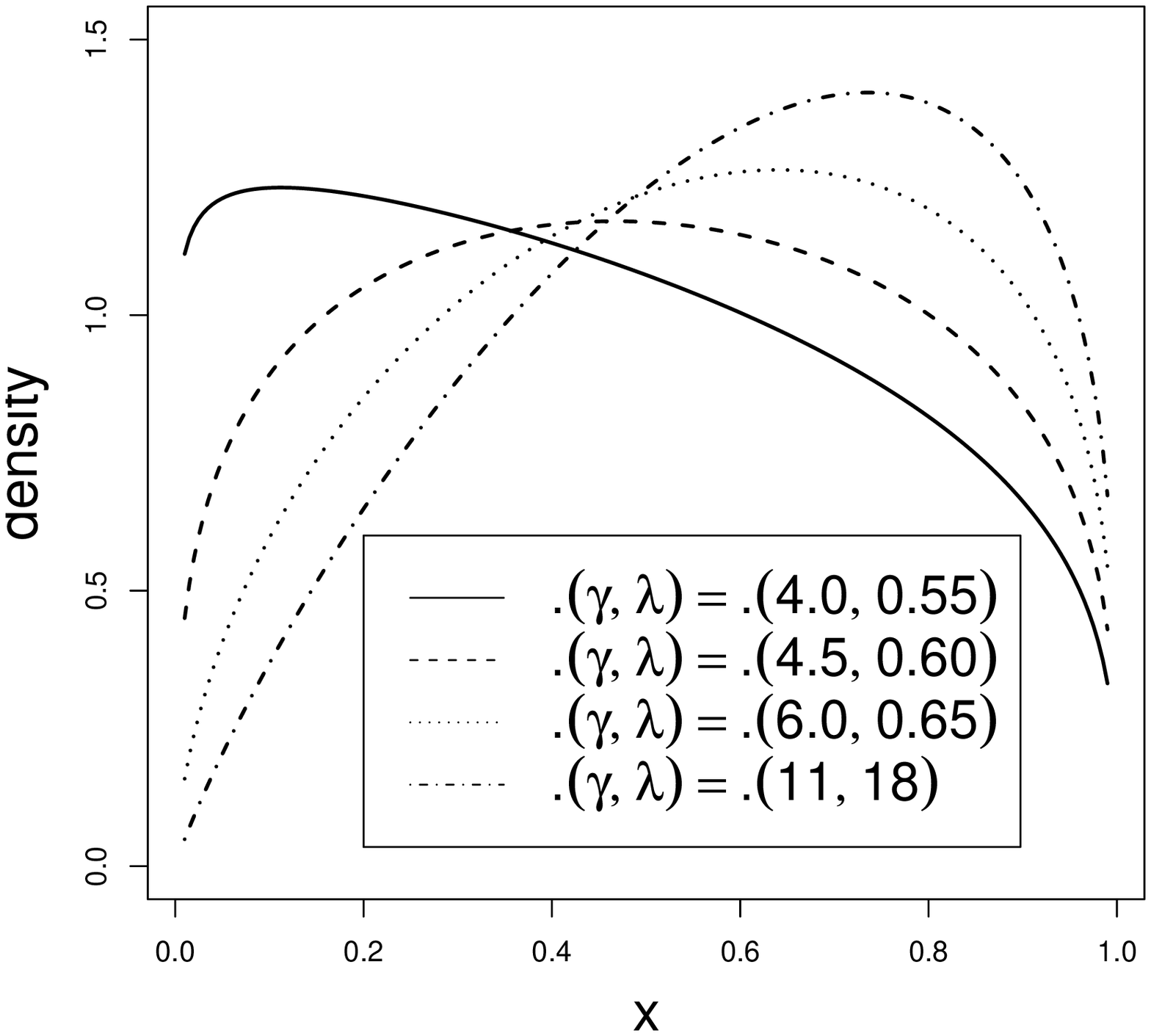}
\begin{center}
(i)
\end{center}
\end{minipage}
\caption{\textit{GKw} density curves.
(a) $\tetn=(\alpha,3.5,1.5,2.5,0.5)^{\top}$,
(b) $\tetn=(3.5,\beta,1.5,2.5,0.5)^{\top}$,
(c) $\tetn=(1.0,1.5,\gamma,2.5,0.5)^{\top}$,
(d) $\tetn=(1.0,1.5,2.5,\delta,0.5)^{\top}$,
(e) $\tetn=(0.5,0.7,0.1,3.0,\lambda)^{\top}$,
(f) $\tetn=(\alpha,\beta,2.5,0.1,0.5)^{\top}$,
(g) $\tetn=(\alpha,1.5,2.5,\delta,0.5)^{\top}$,
(h) $\tetn=(0.5,0.7,0.15,\delta,\lambda)^{\top}$,
(i) $\tetn=(0.5,1.0,\gamma,0.3,\lambda)^{\top}$.
\label{graf_density}}
\end{figure}

Immediately, inverting the transformation motivation (\ref{FGKw1}), we can generate
$X$ following the \textit{GKw} distribution by $X=[1-(1-V^{1/\lambda})^{1/\beta}]^{1/\alpha}$,
where $V$ is a beta random variable with parameters $\gamma$ and $\delta+1$.
This scheme is useful because of the existence of fast generators for beta
random variables. Figure \ref{graf_density} plots some of the possible shapes of the density
function (\ref{fGKw}). The \textit{GKw} density function can take various forms,
bathtub, $J$, inverted $J$, monotonically increasing or decreasing and upside-down bathtub,
depending on the parameter values.

\vspace{0.5cm}

We now provide two properties of the \textit{GKw} distribution.
\vspace{0.5cm}

\noindent{\bf Proposition 1.}
If $X \sim {\rm GKw}(1,\beta,\gamma,\delta,\lambda)$, then $Y=X^{1/\alpha}\sim {\rm GKw}({\alpha},\beta,\gamma,\delta,\lambda)$ for $\alpha>0$.

\vspace{0.5cm}
\noindent{\bf Proposition 2.}
Let $X \sim {\rm GKw}(\alpha,\beta,\gamma,\delta,\lambda)$ and $Y=-\log(X)$. Then, the pdf of $Y$ is given by
\begin{eqnarray}\label{EGKw}
f(y;\tetn)=\frac{\lambda \alpha \beta}{B(\gamma,\delta+1)}e^{-\alpha y}(1-e^{-\alpha y})^{\beta-1}
[1-(1-e^{-\alpha y})^{\beta}]^{\gamma \lambda -1}\{1-[1-(1-e^{-\alpha y})^{\beta}]^{\lambda}\}^{\delta},\,\,y>0.
\end{eqnarray}
We call (\ref{EGKw}) the log-generalized Kumaraswamy (\textit{LGKw}) distribution.

\section{Special Sub-Models}\label{special distri}

The \textit{GKw} distribution is very flexible and has the following distributions as
special sub-models.

\subsection*{The Kumaraswamy distribution (\textit{Kw})}

If $\lambda=\gamma=1$ and $\delta=0$, the \textit{GKw} distribution reduces to
the \textit{Kw} distribution with parameters $\alpha$ and $\beta$, and cdf and
pdf given by (\ref{G1}) and (\ref{g1}), respectively.

\subsection*{The McDonald distribution (\textit{Mc})}

For $\alpha=\beta=1$, we obtain the \textit{Mc} distribution (\ref{g2})
with parameters $\gamma$, $\delta+1$ and $\lambda$.

\subsection*{The beta distribution}

If $\alpha=\beta=\lambda=1$, the \textit{GKw} distribution reduces to the beta distribution
with parameters $\gamma$ and $\delta+1$.

\subsection*{The beta Kumaraswamy distribution (\textit{BKw})}

If $\lambda=1$, (\ref{fGKw}) yields
\begin{eqnarray*}\label{BKw}
f(x;\alpha,\beta,\gamma,\delta,1)=\frac{\alpha \ \beta \ x^{\alpha-1}}{B(\gamma,\delta+1)}(1-x^{\alpha})^{\beta(\delta+1)-1}
[1-(1-x^{\alpha})^{\beta}]^{\gamma-1},\,\,0<x<1.
\end{eqnarray*}
This distribution can be viewed as a four-parameter generalization of the
\textit{Kw} distribution. We refer to it as the \textit{BKw} distribution since its pdf can be
obtained from (\ref{f}) by setting $G_1(x;\omegn)$ to be the ${\rm Kw}(\alpha,\beta)$
cdf and $g_2(x;\taun)$ to the $\mathfrak{B}(\gamma,\delta+1)$ density function.

\subsection*{The Kumaraswamy-Kumaraswamy distribution (\textit{KwKw}) }

For $\gamma=1$, (\ref{fGKw}) reduces to (for $0<x<1$)
\begin{eqnarray*}
f(x;\alpha,\beta,\gamma,\delta,\lambda)=\lambda \alpha \beta (\delta+1)x^{\alpha-1}(1-x^{\alpha})^{\beta-1}[1-(1-x^{\alpha})^{\beta}]^{\lambda-1}
\{1-[1-(1-x^{\alpha})^{\beta}]^{\lambda}\}^{\delta}.
\end{eqnarray*}
Again, this distribution is a four-parameter generalization of the
\textit{Kw} distribution. It can be obtained from (\ref{f}) by replacing
$G_1(x;\omegn)$ by the cdf of the ${\rm Kw}(\alpha,\beta)$ distribution and
$g_2(x;\taun)$ by the pdf of the ${\rm Kw}(\gamma,\delta+1)$ distribution.
Its cdf has a closed form given by
$$F(x;\alpha,\beta,1,\delta,\lambda)=1-\{1-[1-(1-x^\alpha)^\beta]^\lambda\}^{\delta+1}.$$

\subsection*{The \textit{EKw} distribution}

If $\delta = 0$ and $\gamma=1$, (\ref{fGKw}) gives
\begin{eqnarray*}
f(x;\alpha,\beta,1,0,\lambda)&=&\lambda \alpha \beta x^{\alpha-1} (1-x^{\alpha})^{\beta-1}
[1-(1-x^{\alpha})^{\beta}]^{\lambda-1},\,\,0<x<1.
\end{eqnarray*}
It can be easily seen that the associated cdf can be written as
\begin{eqnarray*}
F(x;\alpha,\beta,1,0,\lambda)=G_1(x;\alpha,\beta)^\lambda,
\end{eqnarray*}
where $G_1(x;\alpha,\beta)$ is the cdf of the ${\rm Kw}(\alpha,\beta)$ distribution. This
distribution was defined before as the \textit{EKw} distribution which is a new three-parameter
generalization of the \textit{Kw} distribution.

\subsection*{The beta power distribution (\textit{BP})}

For $\alpha=1$ and $\beta=1$, (\ref{EGKw}) reduces to
\begin{eqnarray*}
f(x;1,1,\gamma,\delta,\lambda)&=&\frac{\lambda}{B(\gamma,\delta+1)}x^{\gamma \lambda-1}(1-x^{\lambda})^{\delta}, \ \ 0<x<1.
\end{eqnarray*}
This density function can be obtained from (\ref{f}) if $G_1(x)=x^{\lambda}$
and $g_2(x)$ is taken as the beta density with parameters $\gamma$
and $\delta+1$. We call this distribution as the \textit{BP} distribution.

\vspace{0.7cm}
The \textit{LGKw} distribution (\ref{EGKw}) contains as special sub-models
the following distributions.

\subsection*{The beta generalized exponential distribution (\textit{BGE})}

For $\lambda=1$, (\ref{EGKw}) gives
\begin{eqnarray}
\label{bge}
f(y;\alpha,1,\gamma,\delta,\lambda)=\frac{\alpha \beta}{B(\gamma,\delta+1)}e^{-\alpha y}(1-e^{-\alpha y})^{\beta(\delta+1)-1}[1-(1-e^{-\alpha y})^{\beta}]^{\gamma-1},\,\,y>0,
\end{eqnarray}
which is the density function of the \textit{BGE} distribution introduced by
Barreto-Souza et al.~(2010). If $\gamma=1$ and $\delta = 0$ in addition to $\lambda=1$,
the \textit{LGKw} distribution becomes the generalized exponential distribution (Gupta and Kundu, 1999).
If $\lambda=\beta=\gamma=1$ and $\delta=0$, (\ref{bge}) coincides with the exponential
distribution with mean $1/\alpha$.

\subsection*{The beta exponential distribution (\textit{BE})}

For $\beta=1$ and $\lambda=1$, (\ref{EGKw}) reduces to
\begin{eqnarray*}
f(y;\alpha,1,\gamma,\delta,1)=\frac{\alpha}{B(\gamma,\delta+1)}e^{-\alpha \gamma y}(1-e^{-\alpha y})^{\delta},\,\,y>0,
\end{eqnarray*}
which is the density of the \textit{BE} distribution introduced by Nadarajah and Kotz~(2006).

\section{Expansions for the Distribution and Density Functions}

We now give simple expansions for the cdf of the {\it GKw} distribution. If $|z|<1$
and $\delta>0$ is a non-integer real number, we have
\begin{eqnarray}
\label{expansion}
(1-z)^{\delta}=\sum_{j=0}^{\infty}(-1)^{j}(\delta)_{j}z^{j},
\end{eqnarray}
where $(\delta)_{j}=\delta(\delta-1)\ldots(\delta-j+1)$ (for $j=0,1,\dots$) is the
descending factorial. Clearly, if $\delta$ is a positive integer, the series stops
at $j=\delta$. Using the series expansion (\ref{expansion}) and the representation
for the {\it GKw} cdf (\ref{FGKw}), we obtain
\begin{eqnarray*}
F(x;\tetn)=\int_{0}^{G_{1}(x;\alpha,\beta)}\frac{\lambda}{B(\gamma,\delta+1)}y^{\gamma \lambda-1}\sum_{j=0}^{\infty}(\delta)_{j}(-1)^{j}y^{\lambda j}dy
\end{eqnarray*}
if $\delta$ is a non-integer real number. By simple integration, we have
\begin{eqnarray}\label{Fweight}
F(x;\tetn)=\sum_{j=0}^{\infty}\omega_{j}[G_{1}(x;\alpha,\beta)]^{\lambda(\gamma+j)},
\end{eqnarray}
where
\begin{eqnarray}
\label{weight}
\omega_{j}=\frac{(-1)^{j}(\delta)_{j}}{(\gamma+j)B(\gamma,\delta+1)},
\end{eqnarray}
and $G_{1}(x;\alpha,\beta)$ is given by (\ref{G1}). If $\delta$ is a positive integer,
the sum stops at $j=\delta$.

The moments of the \textit{GKw} distribution do not have closed form. In order to
obtain expansions for these moments, it is convenient to develop expansions
for its density function. From (\ref{Fweight}), we can write
\begin{eqnarray*}
f(x;\tetn)=\sum_{j=0}^{\infty}\omega_{j}\lambda(\gamma+j)g_{1}(x;\alpha,\beta)[G_{1}(x;\alpha,\beta)]^{\lambda(\gamma+j)-1}.
\end{eqnarray*}
If we replace $G_{1}(x;\alpha,\beta)$ by (\ref{G1}) and use (\ref{g1}), we obtain
\begin{eqnarray}\label{fmistura}
f(x;\tetn)=\sum_{k=0}^{\infty}p_k\,g_{1}(x;\alpha,(k+1)\beta),
\end{eqnarray}
where $p_k=\sum_{j=0}^{\infty}\omega_{j}\,t_{j,k},$
with $t_{j,k}=(\phi)_{k}\,\lambda (\gamma+j)(-1)^{k}/(k+1).$
Here, $\phi=(\gamma+j)\lambda-1$ and $g_{1}(x;\alpha,(k+1)\beta)$ denotes
the ${\rm Kw}(\alpha,(k+1)\beta)$ density function with parameters
$\alpha$ and $(k+1)\beta$. Further, we can express (\ref{fmistura})
as a mixture of power densities, since the \textit{Kw} density (\ref{g1}) can
also be written as a mixture of power densities. After some algebra, we obtain
\begin{eqnarray}\label{mispower}
f(x;\tetn)=\sum_{i=0}^{\infty}v_i\,x^{(i+1)\alpha-1},
\end{eqnarray}
where $$v_i=(-1)^i\,\alpha\,\beta\,\sum_{k=0}^{\infty}(k+1)\,((k+1)\beta-1)_i\,p_k.$$
Equations (\ref{fmistura}) and (\ref{mispower}) are the main results of this
section. They can provide some mathematical properties of the \textit{GKw}
distribution from the properties of the \textit{Kw} and power
distributions, respectively.

\section{Moments and Moment Generating Function}

Let $X$ be a random variable having the {\it GKw} distribution (\ref{fGKw}).
First, we obtain an infinite sum representation for the \textit{rth} ordinary moment
of $X$, say $\mu_{r}^{\prime}=E(X^r)$. From (\ref{fmistura}), we can write
\begin{eqnarray}
\label{momentos}
\mu_{r}^{\prime}=\sum_{k=0}^{\infty}p_k\,\tau_{r}(k),
\end{eqnarray}
where $\tau_{r}(k)=\int_{0}^{1}x^{r}g_{1}(x;\alpha,(k+1)\beta) dx$ is
the \textit{rth} moment of the ${\rm Kw}(\alpha,(k+1)\beta)$
distribution which exists for all $r>-\alpha$. Using a result due to
Jones (2009, Section 3), we have
\begin{eqnarray}\label{momentos_Kw}
\tau_{r}(k)=(k+1)\beta B\Big(1+\frac{r}{\alpha},(k+1)\beta\Big).
\end{eqnarray}
Hence, the moments of the {\it GKw} distribution follow directly from
(\ref{momentos}) and (\ref{momentos_Kw}). The central moments $(\mu_s)$ and cumulants $(\kappa_s)$ of $X$ are easily obtained
from the ordinary moments by
$\mu_s=\sum_{k=0}^{s} \binom{s}{k}(-1)^k \mu_1^{\prime s}\mu_{s-k}^{\prime}$ and
$\kappa_1=\mu^{\prime}_1,\,\kappa_2=\mu^{\prime}_2-\mu^{\prime2}_1,\,
\kappa_3=\mu^{\prime}_3-3\mu^{\prime}_2\mu^{\prime}_1+2\mu^{\prime3}_1,\,
\kappa_4=\mu^{\prime}_4-4\mu^{\prime}_3\mu^{\prime}_1 -
3\mu^{\prime2}_2+12\mu^{\prime}_2\mu^{\prime 2}_1
-6\mu^{\prime 4}_1,\ \kappa_5=\mu^{\prime}_5 -
5\mu^{\prime}_4\mu^{\prime}_1-10\mu^{\prime}_3\mu^{\prime}_2+
20\mu^{\prime}_3\mu^{\prime 2}_1+30\mu^{\prime 2}_2\mu^{\prime}_1-
60\mu^{\prime}_2\mu^{\prime 3}_1+24\mu^{\prime 5}_1,
\kappa_6 =
\mu^{\prime}_6-6\mu^{\prime}_5\mu^{\prime}_1-15\mu^{\prime}_4\mu^{\prime}_2
+ 30\mu^{\prime}_4 \mu^{\prime2}_1 - 10
\mu^{\prime2}_3+120\mu^{\prime}_3\mu^{\prime}_2\mu^{\prime}_1-
120\mu^{\prime}_3\mu^{\prime3}_1+30\mu^{\prime 3}_2
-270\mu^{\prime 2}_2\mu^{\prime 2}_1+360\mu^{\prime}_2\mu^{\prime
  4}_1-120\mu^{\prime6}_1$,
etc., respectively. The $r$th descending factorial moment of $X$ is
$$\mu_{(r)}^{\prime}=\text{E}[X^{(r)}]=\text{E}[X(X-1)\times \cdots
\times (X-r+1)]=\sum_{m=0}^{r} s(r,m)\,\mu^{\prime}_m,$$
where $s(r,m)$ is the Stirling number of the first kind given by
$s(r,m)=(m!)^{-1} {d^m}x^{(r)}/{d x^m}|_{x=0}$. It counts
the number of ways to permute a list of $r$ items into $m$ cycles. Thus,
the factorial moments of $X$ are given by
\begin{equation*}
\mu_{(r)}^{\prime}=\sum_{k=0}^{\infty}p_k \sum_{m=0}^{r} s(r,m)\,\tau_{m}(k).
\end{equation*}

The moment generating function of the {\it GKw} distribution, say $M(t)$, is obtained from (\ref{mispower})
as
\begin{eqnarray*}
M(t)=\sum_{i=0}^{\infty}v_i\,\int_{0}^{1} x^{(i+1)\alpha-1} \exp(t x) dx.
\end{eqnarray*}
By changing variable, we have
\begin{eqnarray*}
M(t)=\sum_{i=0}^{\infty}v_i\, t^{-(i+1)\alpha}\,\int_{0}^{t}\,u^{(i+1)\alpha-1}\,\exp(-u)du
\end{eqnarray*}
and then $M(t)$ reduces to the linear combination
\begin{eqnarray*}
M(t)=\sum_{i=0}^{\infty}v_i\,\frac{\gamma((i+1)\alpha,t)}{t^{(i+1)\alpha}},
\end{eqnarray*}
where $\gamma(a,x)=\int_{0}^{a}\,u^{a-1}\,\exp(-u)du$ denotes the incomplete
gamma function.

\section{Quantile Function}

We can write (\ref{FGKw1}) as $F(x;\tetn)=I_{z}(\gamma,\delta+1)=u$, where
$z=[1-(1-x^\alpha)^{\beta}]^{\lambda}$.
From Wolfram's website\footnote{http://functions.wolfram.com/06.23.06.0004.01}
we can obtain some expansions for the inverse of the 
incomplete beta function, say $z=Q_{\it B}(u)$, 
one of which is
\begin{eqnarray*}
z=Q_{\it B}(u)=a_1 v+a_2 v^2+a_3 v^3+a_4v^4+O(v^{5/\gamma}),
\end{eqnarray*}
where $v =[\gamma u B(\gamma,\delta+1)]^{1/\gamma}$ for $\gamma>0$ and
$a_0=0$, $a_1=1$, $a_2=\delta/(\gamma+1)$,
$$a_3= \frac{\delta[\gamma^2+3(\delta+1)\gamma-
\gamma + 5\delta+1]}{2(\gamma+1)^2(\gamma+2)},$$
\begin{eqnarray*}
a_4&=&\delta\{\gamma^4+(6\delta+5)\gamma^3 + (\delta+3)(8\delta+3)\gamma^2
+[33(\delta+1)^2-30\delta+26]\gamma\nonumber\\
&+&(\delta+1)(31\delta-16)+18\}/[3(\gamma+1)^3(\gamma+2)(\gamma+3)],\ldots
\end{eqnarray*}
The coefficients $a_i'$s for $i\ge2$ can be derived from the cubic
recursion (Steinbrecher and Shaw, 2007)
\begin{eqnarray*}
a_{i}&=&\frac{1}{[i^2+(\gamma-2)i+(1-\gamma)]}\bigg\{(1-\rho_{i,2})
\sum_{r=2}^{i-1} a_{r}\,a_{i+1-r}\,[r(1-\gamma)(i-r)\nonumber\\
&-&r(r-1)]+\sum_{r=1}^{i-1} \sum_{s=1}^{i-r}a_{r}\,a_{s}\,a_{i+1-r-s}\,[r(r-\gamma)+s(\gamma+\beta-2)
\nonumber\\
&\times&(i+1-r-s)]\bigg\},
\end{eqnarray*}
where $\rho_{i,2}=1$ if $i=2$ and $\rho_{i,2}=0$ if $i \ne 2$.
In the last equation, we note that the quadratic term only contributes
for $i\ge 3$. Hence, the quantile function $Q_{\it GKw}(u)$ of the {\it GKw} distribution
can be written as $Q_{\it GKw}(u)=\{1-[1-Q_B(u)^{1/\lambda}]^{1/\beta}\}^{1/\alpha}$.

\section{Mean Deviations}

If $X$ has the {\it GKw} distribution,
we can derive the mean deviations about the mean $\mu^{\prime}_1=E(X)$ and about
the median $M$ from
\begin{equation*}
\delta_1=\int_{0}^{1}\mid\!x - \mu^{\prime}_1\!\!\mid f(x;\tetn) dx\,\,\,\,\text{and}\,\,\,\,
\delta_2=\int_{0}^{1}\mid\! x - M\!\!\mid f(x;\tetn) dx,
\end{equation*}
respectively. From (\ref{FGKw1}), the median $M$ is the solution of the nonlinear
equation $$I_{[1-(1-M^\alpha)^{\beta}]^{\lambda}}(\gamma,\delta+1)=1/2.$$

These measures can be calculated using the relationships
\begin{equation*}
\delta_1= 2\big[\mu^{\prime}_1 F(\mu^{\prime}_1;\tetn)-J(\mu^{\prime}_1;\tetn)\big]
\,\,\,\,\text{and}\,\,\,\,\delta_2= \mu^{\prime}_1-2 J(M;\tetn).
\end{equation*}
Here, the integral $J(a;\tetn)=\int_{0}^{a} x f(x;\tetn) dx$ is easily calculated from the
density expansion (\ref{mispower}) as
\begin{eqnarray*}
J(a;\tetn)=\sum_{i=0}^{\infty}\frac{v_i\,a^{(i+1)\alpha+1}}{(i+1)\alpha+1}.
\end{eqnarray*}

We can use this result to obtain the Bonferroni and Lorenz curves. These curves have
applications not only in economics to study income and poverty, but also in other fields,
such as reliability, demography, insurance and medicine. They are defined by
\begin{eqnarray*}
\displaystyle
B(p;\tetn)=\frac{J(q;\tetn)}{p \mu^{\prime}_1}\,\,\,\,
\text{and}\,\,\,\,
L(p;\tetn)=\frac{J(q;\tetn)}{\mu^{\prime}_1},
\end{eqnarray*}
respectively, where $q=F^{-1}(p;\tetn)$.

\section{Moments of Order Statistics}

The density function of the \textit{ith} order statistic $X_{i:n}$, say $f_{i:n}(x;\tetn)$,
in a random sample of size $n$ from the {\it GKw} distribution, is given by
(for $i=1,\cdots,n$)
\begin{eqnarray}
\label{f_momento}
f_{i:n}(x;\tetn)=\frac{1}{B(i,n-i+1)}f(x;\tetn)F(x;\tetn)^{i-1}\{1-F(x;\tetn)\}^{n-1},\,\,0<x<1.
\end{eqnarray}
The binomial expansion yields
\begin{eqnarray*}
f_{i:n}(x;\tetn)=\frac{1}{B(i,n-i+1)}f(x;\tetn)\,\sum_{j=0}^{n-1} \binom{n-1}{j}(-1)^j\, F(x;\tetn)^{i+j-1},
\end{eqnarray*}
and using and integrating (\ref{mispower}) we arrive at
\begin{eqnarray*}
f_{i:n}(x;\tetn)=\frac{1}{B(i,n-i+1)}\,\left(\sum_{t=0}^{\infty}v_t\,x^{(t+1)\alpha-1}\right)\sum_{j=0}^{n-1} \binom{n-1}{j}(-1)^j\,\left(\sum_{s=0}^{\infty}v_s^{\star}\,x^{(s+1)\alpha}\right)^{i+j-1},
\end{eqnarray*}
where $v_s^{\star}=v_s[(s+1)\alpha]^{-1}$.

We use the following expansion for a power series raised to a integer power 
(Gradshteyn and Ryzhik, 2000, Section 0.314)
\begin{eqnarray}
\label{serie}
\bigg(\sum_{j=0}^{\infty}a_{j}x^{j}\bigg)^{p}=\sum_{j=0}^{\infty}c_{j,p}x^{j},
\end{eqnarray}
where $p$ is any positive integer number, $c_{0,p}=a_{0}^{p}$ and $c_{s,p}=(sa_{0})^{-1}\sum_{j=1}^{s}(jp-s+j)a_{j}c_{s-j,p}$ for all $s \geq 1$.]
We can write
\begin{eqnarray*}
f_{i:n}(x;\tetn)=\frac{1}{B(i,n-i+1)}\,\sum_{j=0}^{n-1} \binom{n-1}{j}(-1)^j\,\sum_{s,t=0}^{\infty}v_t\,e_{s,i+j-1}\,x^{(s+t+i+j)\alpha-1},
\end{eqnarray*}
where $e_{0,i+j-1}=v_{0}^{\star(i+j-1)}$ and (for $s \ge 1$)
$$e_{s,i+j-1}=(s v_{0}^{\star})^{-1}\sum_{m=1}^{s}[m(i+j-1)-s+m]v_{m}^{\star}e_{s-m,i+j-1}.$$

The \textit{rth} moment of the \textit{ith} order statistic becomes
\begin{eqnarray}\label{order1}
E(X_{i:n}^{r})=\frac{1}{B(i,n-i+1)}\,\sum_{j=0}^{n-1} \binom{n-1}{j}(-1)^j\,\sum_{s,t=0}^{\infty}\frac{v_t\,e_{s,i+j-1}}{(r+s+t+i+j)\alpha}.
\end{eqnarray}

We now obtain another closed form expression for the moments of the {\it GKw}
order statistics using a general result due to Barakat and Abdelkader (2004)
applied to the independent and identically distributed case. For a distribution with pdf $f(x;\tetn)$ and cdf $F(x;\tetn)$, we can write
\begin{eqnarray*}
E(X_{i:n}^{r})=r\sum_{m=n-i+1}^{n}(-1)^{m-n+i-1}\Big(\begin{tabular}{c} $m-1$ \\ $n-i$ \\ \end{tabular} \Big) \Big(\begin{tabular}{c} $n$ \\ $m$ \\ \end{tabular} \Big)I_{m}(r),
\end{eqnarray*}
where
\begin{eqnarray*}\label{Im}
I_{m}(r)=\int_{0}^{1}x^{r-1}\{1-F(x;\tetn)\}^{m}dx.
\end{eqnarray*}
For a positive integer $m$, we have
\begin{eqnarray*}
I_{m}(r)=\int_{0}^{1}x^{r-1}\sum_{p=0}^{m} \Big(\begin{tabular}{c} $m$ \\ $p$ \\ \end{tabular} \Big)(-1)^{p}[F(x;\tetn)]^{p}dx.
\end{eqnarray*}
By replacing (\ref{Fweight}) in the above equation we have
\begin{eqnarray}
\label{Im2}
I_{m}(r)=\sum_{p=0}^{m}(-1)^{p}\Big(\begin{tabular}{c} $m$ \\ $p$ \\ \end{tabular} \Big)\int_{0}^{1}x^{r-1}\Bigg(\sum_{j=0}^{\infty}\omega_{j}[G_{1}(x;\alpha,\beta)]^{\lambda(\gamma+j)}\Bigg)^{p}dx.
\end{eqnarray}

Equations (\ref{serie}) and (\ref{Im2}) yield
\begin{eqnarray*}
I_{m}(r)=\sum_{p=0}^{m}\Big(\begin{tabular}{c} $m$ \\ $p$ \\ \end{tabular} \Big)(-1)^{p}\int_{0}^{1}x^{r-1}\sum_{j=0}^{\infty}c_{j,p}[G_{1}(x;\alpha,\beta)]^{\lambda(\gamma+j)}dx.
\end{eqnarray*}
By replacing $G_{1}(x;\alpha,\beta)$ by (\ref{G1}) and using (\ref{expansion}) we obtain
\begin{eqnarray*}
I_{m}(r)=\sum_{p=0}^{m}\Big(\begin{tabular}{c} $m$ \\ $p$ \\ \end{tabular} \Big)(-1)^{p}\sum_{j,w=0}^{\infty}(-1)^{w}\,c_{j,p}\,(\psi)_{w}\int_{0}^{1}x^{r-1}(1-x^{\alpha})^{w\beta}dx,
\end{eqnarray*}
where $\psi=\lambda(\gamma+j)$. Since $B(a/b,c)=b\int_{0}^{1}w^{a-1}(1-w^{b})^{c-1} dw$
for $a,b,c>0$ (Gupta and Nadarajah, 2004b), we have
\begin{eqnarray*}
I_{m}(r)=\sum_{p=0}^{m}\sum_{j,w=0}^{\infty}s_{p,j,w}\,B(\frac{r}{\alpha},\beta w+1),
\end{eqnarray*}
where
\begin{eqnarray*}
s_{p,j,w}=\frac{(-1)^{p+w}m!}{\alpha(m-p)!p!}c_{j,p}(\psi)_{w}.
\end{eqnarray*}
Finally, $E(X_{i:n}^{r})$ reduces to 
\begin{eqnarray}\label{momorder}
E(X_{i:n}^{r})= r\sum_{m=n-i+1}^{n}\Bigg\{(-1)^{m-n+i-1}\Big(\begin{tabular}{c} $m-1$ \\ $n-i$ \\ \end{tabular} \Big) \Big(\begin{tabular}{c} $n$ \\ $m$ \\ \end{tabular} \Big) \sum_{p=0}^{m}\sum_{j,w=0}^{\infty}s_{p,j,w}B\left(\frac{r}{\alpha},\beta w+1\right) \Bigg\}.
\end{eqnarray}

Equations (\ref{order1}) and (\ref{momorder}) are the main results of this section. 
The L-moments are analogous to the ordinary moments but can be estimated by linear
combinations of order statistics. They are linear functions of expected order
statistics defined by (Hoskings, 1990)
$$\lambda_{r+1}=(r+1)^{-1} \sum_{k=0}^{r} (-1)^k \binom{r}{k}
E(X_{r+1-k:r+1}),\,\,r=0,1,\ldots$$
The first four L-moments are
$\lambda_1=E(X_{1:1})$, $\lambda_2=\frac{1}{2}E(X_{2:2}-X_{1:2})$,
$\lambda_3=\frac{1}{3}E(X_{3:3}-2X_{2:3}+X_{1:3})$ and
$\lambda_4=\frac{1}{4}E(X_{4:4}-3X_{3:4}+ 3X_{2:4}-X_{1:4})$.
These moments have several advantages over the ordinary moments. For example, they exist
whenever the mean of the distribution exists, even though some higher moments may not exist,
and are relatively robust to the effects of outliers. From (\ref{momorder}) applied for
the means ($r=1$), we can obtain expansions for the L-moments of the {\it GKw}
distribution.

\section{R\'enyi Entropy}

The entropy of a random variable $X$ with density function $f(x)$ is a measure
of variation of the uncertainty. One of the popular entropy measures is
the R\'enyi entropy given by
\begin{equation}\label{entp01}
\mathcal{J}_{R}(\rho)=\frac{1}{1-\rho}\log\bigg[\int{f^{\rho}(x)} dx \bigg],\,\,\rho > 0,
\,\,\rho \neq 1.
\end{equation}
From (\ref{mispower}), we have
\begin{eqnarray*}
f(x;\tetn)^{\rho}=\left(\sum_{i=0}^{\infty}v_i\,x^{(i+1)\alpha-1}\right)^{\rho}.
\end{eqnarray*}
In order to obtain an expansion for the above power series for $\rho>0$, we can write
\begin{eqnarray*}
f(x;\tetn)^{\rho}&=&
\sum_{j=0}^\infty \binom{\rho}{j}(-1)^j \left\{1-\left(\sum_{i=0}^{\infty}v_i
\,x^{(i+1)\alpha-1}\right)\right\}^j\\
&=&\sum_{j=0}^\infty \sum_{r=0}^j (-1)^{j+r}
\binom{\rho}{j}\binom{j}{r} x^{(\alpha-1)r}\left(\sum_{i=0}^{\infty}v_i\,x^{i\alpha}\right)^r.
\end{eqnarray*}
Using equation (\ref{serie}), we obtain
\begin{eqnarray*}
f(x;\tetn)^{\rho}=\sum_{i,j=0}^\infty \sum_{r=0}^j (-1)^{j+r}
\binom{\rho}{j}\binom{j}{r}d_{i,r}\,x^{(i+r)\alpha-r},
\end{eqnarray*}
where $d_{0,r}=v_{0}^{r}$ and $d_{s,r}=(s v_{0})^{-1}\sum_{m=1}^{s}(mr-s+m)v_{m}d_{s-m,r}$ for all $s \geq 1$. Hence,
\begin{eqnarray*}
\mathcal{J}_{R}(\rho)=\frac{1}{1-\rho}\log\bigg[\sum_{i,j=0}^\infty\sum_{r=0}^j \frac{(-1)^{j+r}
\binom{\rho}{j}\binom{j}{r}d_{i,r}}{(i+r)\alpha-r+1}\bigg].
\end{eqnarray*}

\section{Maximum Likelihood Estimation}

Let $X_{1},X_{2},\ldots,X_{n}$ be a random sample from the
${\rm GKw}(\lambda,\alpha,\beta,\gamma,\delta)$ distribution. From (\ref{fGKw}) the
log-likelihood function is easy to derive. It is given by
\begin{eqnarray*}
\label{log_vero}
\ell(\tetn)&=&n\log(\lambda)+n\log(\alpha)+n\log(\beta)-n\log[B(\gamma,\delta+1)]+
(\alpha-1)\sum_{i=1}^{n}\log(x_{i})\nonumber \\ && +(\beta-1)\sum_{i=1}^{n}\log(1-x_{i}^{\alpha})+
(\gamma \lambda-1)\sum_{i=1}^{n}\log[1-(1-x_{i}^{\alpha})^{\beta}] + \nonumber \\ &&
\delta \sum_{i=1}^{n}\log[1-\{1-(1-x_{i}^{\alpha})^{\beta}\}^{\lambda}].
\end{eqnarray*}
By taking the partial derivatives of the log-likelihood function with respect
to $\lambda$, $\alpha$, $\beta$, $\gamma$ and $\delta$, we obtain the components
of the score vector, $U(\tetn)=(U_{\alpha},U_{\beta},U_{\gamma},U_{\delta},U_{\lambda})$. 
They are given by
\begin{eqnarray*}
\label{U1}
U_{\alpha}(\tetn)&=& \frac{n}{\alpha}+\sum_{i=1}^{n}[1-(\beta-1)z_{i}]\log(x_{i})+(\gamma \lambda -1)\sum_{i=1}^{n}\frac{\dot{y}_{i(\alpha)}}{y_{i}}-\delta \lambda\sum_{i=1}^{n}v_{i}\dot{y}_{i(\alpha)},\\
U_{\beta}(\tetn)&=&\frac{n}{\beta}+\sum_{i=1}^{n}\log(1-x_{i}^{\alpha})+(\gamma \lambda-1)\sum_{i=1}^{n}\frac{\dot{y}_{i(\beta)}}{y_{i}}-\lambda \delta \sum_{i=1}^{n}v_{i}\dot{y}_{i(\beta)},\\
U_{\gamma}(\tetn)&=&-n[\psi(\gamma)-\psi(\gamma+\delta+1)]+\lambda\sum_{i=1}^{n}\log(y_{i}),\\
U_{\delta}(\tetn)&=&-n[\psi(\delta+1)-\psi(\gamma+\delta+1)]+\sum_{i=1}^{n}\log(1-y_{i}^{\lambda}),\\
U_{\lambda}(\tetn)&=&\frac{n}{\lambda}+\sum_{i=1}^{n}[\gamma-\delta y_{i} v_{i}]\log(y_{i}),
\end{eqnarray*}
where $\psi(\cdot)$ is the digamma function, $y_{i}=1-(1-x^{\alpha}_{i})^{\beta}$, $v_{i}=y^{\lambda-1}_{i}(1-y_{i}^{-\lambda})^{-1}$, $z_{i}=x_{i}^{\alpha}(1-x_{i}^{\alpha})^{-1}$, $\dot{y}_{i(\alpha)}=\partial y_{i}/ \partial \alpha=-\beta x_{i}^{\alpha}(1-x_{i}^{\alpha})^{\beta-1}\log(x_{i})$ and $\dot{y}_{i(\beta)}=\partial y_{i}/ \partial \beta=-(1-x_{i}^{\alpha})^{\beta}\log(1-x_{i}^{\alpha})$.For interval estimation and hypothesis tests on the model parameters, the observed
information matrix is required. The observed information matrix $J=J(\tetn)$ is given in the Appendix.

Under conditions that are fulfilled for parameters in the interior of the parameter
space, the approximate distribution of
$\sqrt{n}(\widehat{\tetn}-\tetn)$ is multivariate normal $N_{5}(\textbf{0},I^{-1}(\tetn))$,
where $\widehat{\tetn}$ is the maximum likelihood estimator (MLE) of $\tetn$ and
$I(\tetn)$ is the expected information matrix. This approximation is also valid
if $I(\tetn)$ is replaced by $J(\widehat{\tetn})$.

The multivariate normal $N_{5}(\textbf{0},J^{-1}(\widehat{\tetn}))$ distribution
can be used to construct approximate confidence regions.
The well-known likelihood ratio (\textit{LR}) statistic can be used for testing
hypotheses on the model parameters in the usual way. In particular, this statistic is
useful to check if the fit using the \textit{GKw}
distribution is statistically superior to a fit using the \textit{BKw}, \textit{EKw}
and \text{Kw} distributions for a given data set. For example, the test of $H_{0}:\lambda=1$
versus $H_{1}: \lambda \neq 1$ is equivalent to compare the \textit{BKw} distribution
with the \textit{GKw} distribution and the \textit{LR} statistic reduces to
$w=2[\ell(\widehat{\alpha},\widehat{\beta},\widehat{\gamma},\widehat{\delta},\widehat{\lambda})-\ell(\tilde{\alpha},\tilde{\beta},\tilde{\gamma},\tilde{\delta},1)]$, where $\widehat\tetn$ and $\widetilde \tetn$ are the unrestricted and restricted MLEs
of $\tetn$, respectively. Under the null hypothesis, $w$ is asymptotically distributed as $\chi^{2}_{1}$. For a given level $\zeta$, the \textit{LR} test rejects $H_{0}$ if $w$
exceeds the $(1-\zeta)$-quantile of the $\chi_{1}^{2}$ distribution.

\section{Application}

\begin{figure} [!htb]
\centering \scalebox{.9}{\includegraphics*{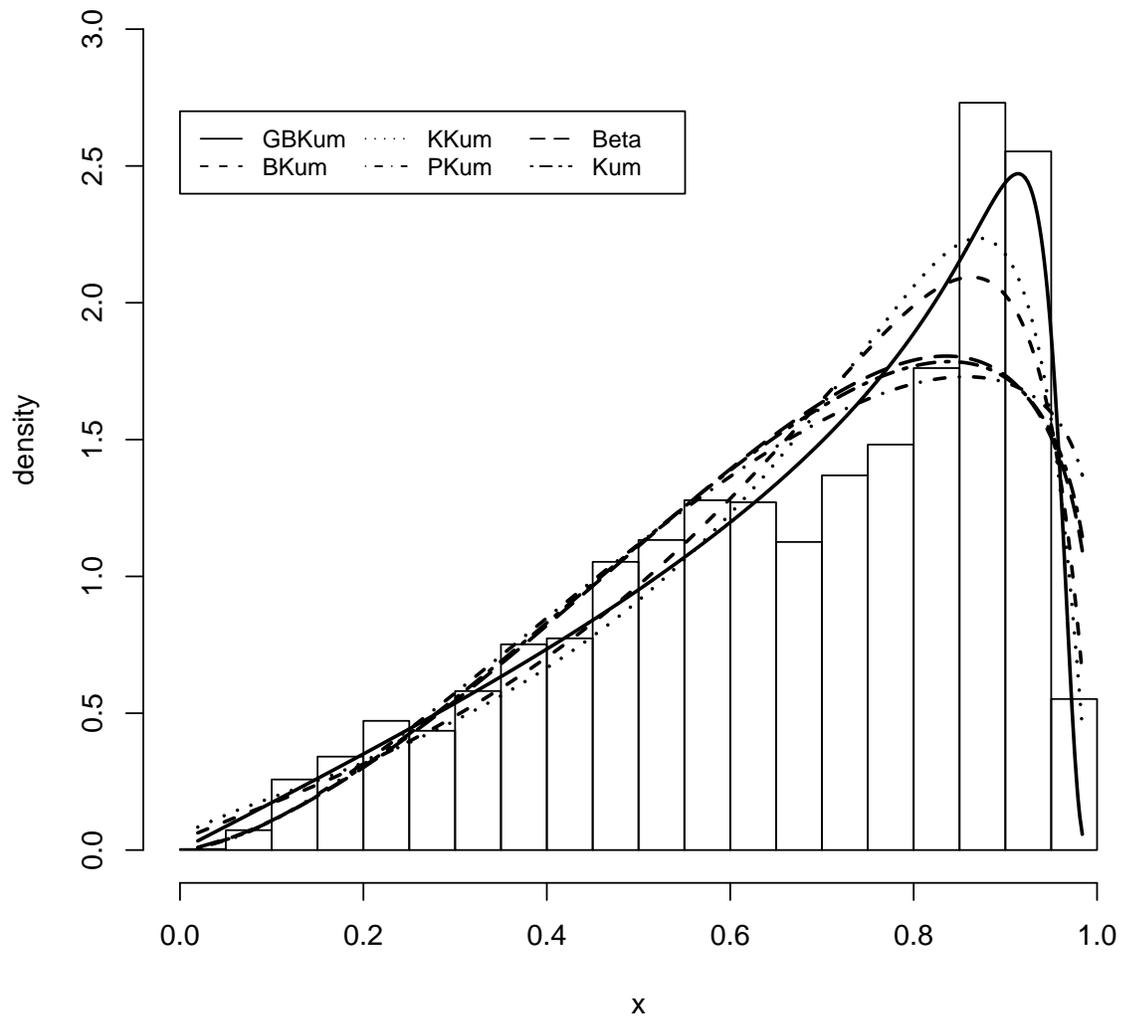}}
\vspace{-0.2cm}
\caption{Histogram and estimated pdf's for the percentage of children living in households
with per capita income less than R\$ 75.50 (1991) in 5509 Brazilian municipal
districts.
\label{figura1}}
\end{figure}

This section contains an application of the \textit{GKw} distribution to real data. The data are the observed percentage of children living in households with per capita income less than R\$ 75.50 in 1991 in 5509 Brazilian municipal districts. The data were extracted from the Atlas of Brazil Human Development database available at http://www.pnud.org.br/. The histogram of the data is shown
in Figure \ref{figura1} along with the estimated densities of the {\it GKw} distribution
and some special sub-models. Apparently, the {\it GKw} distribution gives the best fit.

The \textit{GKw} model includes some sub-models described in Section \ref{special distri} as especial cases and thus allows their evaluation relative to each other and to a more general model. As mentioned before, we can compute the maximum values of the unrestricted and restricted log-likelihoods to obtain the \textit{LR} statistics for testing some sub-models of the \textit{GKw} distribution.
We test $H_{0}:(\alpha,\beta,\lambda)=(1,1,1)$ versus $H_{1}: H_{0}$ \textrm{is not true},
i.e. we compare the \textit{GKw} model with the beta model. In this case,  $w=2\{1510.7-1271.6)\}=239.1$ (p-value$<0.001$) indicates that the
\textit{GKw} model gives a better representation of the data than
the beta distribution. Further, the \textit{LR} statistic for testing 
 $H_{0}:\lambda=1$ versus $H_{1}:\lambda\neq 1$, i.e. to
compare the \textit{GKw} model with the \textit{BKw} model, is
$w=2(1510.7-1383.6)=254.2$ (p-value$<0.001$). It also yields favorable indication
for the \textit{GKw} model. Table \ref{EMV} lists the MLEs of the
model parameters (standard errors in parentheses) for different models. The computations were
carried out using the subroutine \verb"MAXBFGS" implemented in the \texttt{Ox} matrix
programming language (Doornik, 2007).

\begin{small}
\begin{table}[!htb]
\centering {\caption{MLEs of the model parameters.}
\vspace*{0.3cm}
\label{EMV}
\begin{tabular}{cccccc|c}
  \hline
  Distribution & $\alpha$ & $\beta$ & $\gamma$ & $\delta$ & $\lambda$ & $\ell(\widehat{\tetn})$ \\
  \hline
  \textit{GKw}  & 18.1161  & 1.8132  & 0.7303   & 0.0609   & 15.7803  & 1510.6670           \\
        & (0.1829)   & (0.0219)  & (0.0057)   & (0.0008)   & (0.8908)   &                    \\
  \textit{BKw}  & 0.0247   & 0.1849  & 26.0933  & 17.3768  &          & 1383.5690           \\
        & (0.0003)   & (0.0005)  & (0.1054)   & (0.0739)   &          &                     \\
  \textit{KKw}  & 2.7191   & 0.4654  &          & 0.0968   & 79.9999  & 1405.0650           \\
        & (0.0086)   & (0.0060)  &          & (0.0005)   & (1.0915)   &                     \\
  \textit{PKw}  & 17.9676  & 0.1647  &          &          & 1.1533   & 1237.5800           \\
        & (0.2421)   & (0.0019)  &          &          & (0.0054)   &                     \\
  \textit{BP}    &          &         & 0.1590   & 16.7313  & 0.2941   & 1269.9760           \\
        &          &         & (0.0018)   & (0.1998)   & (0.0129)   &                     \\
  \textit{Kw}   & 2.4877   & 1.3369  &          &          &          & 1278.7860           \\
        & (0.0295)   & (0.0180)  &          &          &          &                     \\
  \textit{Beta}  &          &         & 2.5678   & 0.3010   &          & 1271.5610           \\
        &          &         & (0.0317)         &   (0.0147)       &          & \\
\hline
\end{tabular}}
\end{table}
\end{small}

\section{Conclusions}

We introduce a new five-parameter continuous distribution on the standard unit
interval which generalizes the beta, Kumaraswamy (Kumaraswamy, 1980)
and McDonald (McDonald, 1984) distributions and includes as special sub-models
other distributions discussed in the literature. We refer to the new model as the generalized
Kumaraswamy distribution and study some of its mathematical properties.
We demonstrate that the generalized Kumaraswamy density function can be expressed as a mixture
of Kumaraswamy and power densities. We provide the moments and a closed form expression
for the moment generating function. Explicit expressions are derived for the mean
deviations, Bonferroni and Lorenz curves and R\'enyi's entropy. The density of
the order statistics can also be expressed in terms of an infinite mixture
of power densities. We obtain two explicit expressions for their moments.
Parameter estimation is approached by maximum likelihood. The usefulness of
the new distribution is illustrated in an analysis of real data.
We hope that the proposed extended model may attract wider applications in the
analysis of proportions data.

\section*{Acknowledgments}
We gratefully acknowledge financial support from FAPESP and CNPq.

\section*{Appendix}
The elements of the observed information matrix $J(\tetn)$ for $(\alpha,\beta,\gamma,\delta,\lambda)$ are
\begin{eqnarray*}
J_{\alpha \alpha}&=& -\frac{n}{\alpha^{2}}-(\beta-1)\sum_{i=1}^{n}\dot{z}_{i(\alpha)}\log(x_{i})+(\gamma \lambda -1)\sum_{i=1}^{n}\bigg\{\frac{\ddot{y}_{i(\alpha)}}{y_{i}}-\bigg(\frac{\dot{y}_{i(\alpha)}}{y_{i}}\bigg)^{2}\bigg\}- \\ && \delta \lambda \sum_{i=1}^{n}(\dot{v}_{i(\alpha)}\dot{y}_{i(\alpha)}+v_{i}\ddot{y}_{i(\alpha)}),\\
J_{\alpha \beta}&=&-\sum_{i=1}^{n}z_{i}\log(x_{i})+(\gamma \lambda-1)\sum_{i=1}^{n}\bigg\{\frac{\ddot{y}_{i(\alpha\beta)}}{y_{i}}-\frac{\dot{y}_{i(\alpha)}\dot{y}_{i(\beta)}}{y_{i}^{2}}\bigg\}-\delta \lambda \sum_{i=1}^{n}(\dot{v}_{i(\beta)}\dot{y}_{i(\alpha)}+v_{i}\ddot{y}_{i(\alpha\beta)}),\\
J_{\alpha \gamma}&=& \lambda \sum_{i=1}^{n}\frac{\dot{y}_{i(\alpha)}}{y_{i}},\ \ \ 
J_{\alpha \delta}= - \lambda \sum_{i=1}^{n}v_{i}\dot{y}_{i(\alpha)},\ \ \ 
J_{\alpha \lambda}= \sum_{i=1}^{n}\{\gamma/y_{i}-\delta v_{i}\}\dot{y}_{i(\alpha)},
\end{eqnarray*}
\begin{eqnarray*}
J_{\beta \beta}&=& -\frac{n}{\beta^{2}}+(\gamma \lambda-1)\sum_{i=1}^{n}\bigg\{\frac{\ddot{y}_{i(\beta)}}{y_{i}}-\bigg(\frac{\dot{y}_{i(\beta)}}{y_{i}}\bigg)^{2}\bigg\}-\delta \lambda \sum_{i=1}^{n}(\dot{v}_{i(\beta)}\dot{y}_{i(\beta)}+v_{i}\ddot{y}_{i(\beta)}),\\
J_{\beta \gamma}&=& \lambda \sum_{i=1}^{n}\frac{\dot{y}_{i(\beta)}}{y_{i}},\ \ \
J_{\beta \delta}= - \lambda \sum_{i=1}^{n}v_{i}\dot{y}_{i(\beta)},\ \ \
J_{\beta \lambda}=\gamma \sum_{i=1}^{n}\frac{\dot{y}_{i(\beta)}}{y_{i}}- \delta \sum_{i=1}^{n}v_{i}\dot{y}_{i(\beta)},\\
J_{\gamma \gamma}&=&-n \{\psi^{'}(\gamma)-\psi^{'}(\gamma+\delta+1)\},\ \ \
J_{\gamma\delta}=n \psi^{'}(\gamma+\delta+1),\ \ \
J_{\gamma\lambda}=\sum_{i=1}^{n}\log(y_{i}),\\
J_{\delta \delta}&=& -n \{\psi^{'}(\delta+1)-\psi^{'}(\gamma+\delta+1)\},\ \ \
J_{\delta \lambda}= -\sum_{i=1}^{n}y_{i}v_{i}\log(y_{i}),\\
J_{\lambda \lambda}&=& -\frac{n}{\lambda^{2}}-\delta\sum_{i=1}^{n}y_{i}\dot{v}_{i(\lambda)}\log(y_{i}),
\end{eqnarray*}
where $\dot{z}_{i(\alpha)}=\partial z_{i}/\partial \alpha=(1+z_{i})z_{i}\log(x_{i})$, $\ddot{y}_{i(\alpha)}=\partial^{2} y_{i}/\partial \alpha^{2}=\{1-(\beta-1)z_{i}\}\dot{y}_{i(\alpha)}\log(x_{i})$, $\ddot{y}_{i(\beta)}=\partial^{2} y_{i}/\partial \beta^{2}=\dot{y}_{i(\beta)}\log(1-x^{\alpha}_{i})$, $\ddot{y}_{i(\alpha \beta)}=\partial^{2} y_{i}/\partial \alpha \partial \beta=\{1/\beta+\log(1+x_{i}^{\alpha})\} \dot{y}_{i(\alpha)}$ , $\dot{v}_{i(\alpha)}=\partial v_{i}/\partial \alpha=\{(\lambda-1)/y_{i}+\lambda v_{i}\}v_{i}\dot{y}_{i(\alpha)}$, $\dot{v}_{i(\beta)}=\partial v_{i}/\partial \beta=\{(\lambda-1)/y_{i}+\lambda v_{i}\}v_{i}\dot{y}_{i(\beta)}$,  $\dot{v}_{i(\lambda)}=\partial v_{i}/\partial \lambda=(1+y_{i}v_{i})v_{i}\log(y_{i})$ and $\psi^{'}(\cdot)$ is the first derivative of the digamma function.

\section*{References}

\noindent Barreto-Souza, W., Santos, A. S., Cordeiro, G. M. (2010). The beta generalized exponential distribution. \textit{Journal of Statistical Computation and Simulation}, {\bf 80}, 159-172.

\vspace{0.4cm}

\noindent Barakat, H., Abdelkader, Y.H. (2004). Computing the moments of order statistics from nonidentical random variables. \textit{Statistical Methods and Applications}, {\bf 13}, 15-26.

\vspace{0.4cm}

\noindent Courard-Hauri, D. (2007). Using Monte Carlo analysis to investigate the relationship between overconsumption and uncertain access to one's personal utility function. \textit{Ecological Economics}, {\bf 64}, 152-162.

\vspace{0.4cm}

\noindent Doornik, J. (2007). \textit{Ox: An Object-Oriented Matrix Programming Language}. London: Timberlake Consultants Press.

\vspace{0.4cm}

\noindent Eugene, N., Lee, C., Famoye, F. (2002). Beta-normal distribution and its applications. \textit{Communications in Statistics - Theory and Methods}, {\bf 31}, 497-512.

\vspace{0.4cm}

\noindent Ferreira, J.T., Steel M. (2006). A constructive representation of univariate skewed distribution. \textit{Journal of the American Statistical Association}, {\bf 101}, 823-829.

\vspace{0.4cm}

\noindent Fletcher, S.C., Ponnambalam, K. (1996). Estimation of reservoir yield and storage distribution using moments analysis. \textit{Journal of Hydrology}, {\bf 182}, 259-275.

\vspace{0.4cm}

\noindent Ganji, A., Ponnambalam, K., Khalili, D., Karamouz, M. (2006). Grain yield reliability analysis with crop water demand uncertainty.
\textit{Stochastic Environmental Research and Risk Assessment}, {\bf 20}, 259-277.

\vspace{0.4cm}

\noindent Gradshteyn, I.S., Ryzhik, I.M. (2000). \textit{Table of Integrals, Series, and Products}. New York: Academic Press.
\vspace{0.4cm}

\noindent Gupta, A.K., Kundu, D. (1999). Generalized exponential distributions. \textit{Australian and New Zealand Journal of Statistics}, {\bf 41}, 173-188.

\vspace{0.4cm}

\noindent Gupta, A.K., Nadarajah, S. (2004a). On the moments of the beta normal distribution. \textit{Communications in Statistics-Theory and Methods}, {\bf 33}, 1-13.

\vspace{0.4cm}

\noindent Gupta, A.K., Nadarajah, S. (2004b). \textit{Handbook of Beta Distribution and its Applications}. New York: Marcel Dekker.

\vspace{0.4cm}

\noindent Hoskings, J.R.M. (1990) L-moments: analysis and estimation of distribution
using linear combinations of order statistics. \textit{Journal of the Royal Statistical
Society B}, {\bf 52}, 105-124.

\vspace{0.4cm}

\noindent Jones, M.C. (2009). Kumaraswamy's distributions: A beta-type distribution with some tractability advantages. \textit{Statistical Methodology}, {\bf 6}, 70-81.

\vspace{0.4cm}

\noindent Kumaraswamy, P. (1980). A generalized probability density function for double bounded random processes. \textit{Journal of Hydrology}, {\bf 46}, 79-88.

\vspace{0.4cm}

\noindent McDonald, J.B. (1984). Some generalized function for the size distributions of income. \textit{Econo\-me\-trica}, {\bf 52}, 647-663.

\vspace{0.4cm}

\noindent Nadarajah, S., Gupta, A.K. (2004). The beta Fréchet distribution. \textit{Far East Journal of Theoretical Statistics}, {\bf 14}, 15-24.

\vspace{0.4cm}

\noindent Nadarajah, S., Kotz, S. (2004). The beta Gumbel distribution. \textit{Mathematical Problems in Engineering}, {\bf 4}, 323-332.

\vspace{0.4cm}

\noindent Nadarajah, S. Kotz, S. (2006). The exponentiated type distributions. \textit{Acta Applicandae Mathematicae}, {\bf 92}, 97-111.

\vspace{0.4cm}

\noindent Sanchez, S., Ancheyta, J., McCaffrey, W.C. (2007). Comparison of probability distribution function for fitting distillation curves of petroleum. \textit{Energy and Fuels}, {\bf 21}, 2955-2963.

\vspace{0.4cm}

\noindent Seifi, A., Ponnambalam, K., Vlach, J. (2000). Maximization of manufacturing yield of systems with arbitrary distributions of component values. \textit{Annals of Operations Research}, {\bf 99}, 373-383.

\vspace{0.4cm}

\noindent Steinbrecher, G., Shaw, W.T. (2007). Quantile Mechanics. Department of Theoretical Physics,
Physics Faculty, University of Craiova. Working Paper.

\vspace{0.4cm}

\noindent Sundar, V., Subbiah, K. (1989). Application of double bounded probability density-function for analysis of ocean waves.
\textit{Ocean Engineering}, {\bf 16}, 193-200.

\vspace{0.4cm}

\noindent Wahed, A.S. (2006). A general method of constructing extended families of distribution from an existing continuous class. \textit{Journal of Probability and Statistical Science}, {\bf 4}, 165-177.

\end{document}